\documentstyle[11pt,verynewpasp,twoside,psfig,epsf]{article}
\def\edcomment#1{\iffalse\marginpar{\raggedright\sl#1\/}\else\relax\fi}
\reversemarginpar

\begin{document}
\title{Search for radio recombination lines towards the gravitational lens 
PKS 1830-211}
\author{Niruj R. Mohan}
\affil{Raman Research Institute, Bangalore 560080, India \&\\
Indian Institute of Science, Bangalore 560012, India}
\author{K.R. Anantharamaiah}
\affil{Raman Research Institute, Bangalore 560080, India}
\author{W.M. Goss}
\affil{NRAO, Socorro, NM 87801, USA}

\begin{abstract}
A search for radio recombination lines near 20 cm at z=0.193 and z=0.886
towards the gravitational lens system PKS 1830-211 has yielded upper
limits of $|\tau_L| \leq$~5 $\times$ 10$^{-5}$~and $\leq$~5 $\times$ 
10$^{-4}$~ at the two redshifts respectively. Based on the 
non-detections, we derive upper
limits to the emission measure of the ionized gas in the absorbing
systems. We also present continuum flux density measurements over the
frequency range 0.3$-$45 GHz made at a single epoch.
\end{abstract}

\vspace{-0.4in}
\section{Introduction}
\vspace{-0.1in}
There is evidence that the gravitational lens system PKS 1830-211
has two absorbers at two different redshifts of 0.886 and 0.193. 
The main lens, a normal galaxy at z=0.886, has been 
studied in HI, OH and a host of molecular lines (Chengalur, de Bruyn, 
\& Narasimha 1999 and references therein).
The only evidence for an additional absorber at
z=0.193 is from HI absorption (Lovell et al. 1996; Verheijen et al. 1999)
studies. Searches for molecular lines at this redshift have been unsuccessful. 

\vspace{-0.6cm}
\section{Observations}
\vspace{-0.4cm}
Taking advantage~~of~~the strong~radio~~continuum of the 
background source (S$_{1.4~GHz}$ =10 Jy),
we searched for stimulated emission recombination lines at both 
redshifts, using the VLA{\footnote {The National Radio Astronomy Observatory is a facility of
the National Science Foundation operated under cooperative agreement
by Associated Universities, Inc.}}
at 20 cm ; 
the H158$\alpha$~line from the z=0.193 system 
($\nu_{rest}$=1.65 GHz) using a bandwidth of 1.56 MHz and a velocity 
resolution of 5.3 km s$^{-1}$ and the H136$\alpha$~line 
from the z=0.886 system ($\nu_{rest}$=2.59 GHz) using a bandwidth of 3.125 MHz
and a resolution of 21 km s$^{-1}$. Neither line was detected 
with 5$\sigma$~upper
limits to $|\tau_L|$~of 5 $\times$~10$^{-5}$~and 5 $\times$ 10$^{-4}$ 
corresponding to z=0.193 and z=0.886 respectively.

Since this source is highly variable, we used the VLA to measure the
continuum flux density at a single epoch over the frequency range 
327 MHz to 45 GHz, to determine the intrinsic spectrum of the background source.
The observations were made on 27 Oct, 1997 with 3C 286
as the primary calibrator. Figure 1 (right) shows the continuum flux
density as a function of frequency. Assuming that the core radiation 
at low frequencies has a flat spectrum, we derive upper limits to the free-free 
optical depth of $\tau_c$ $\leq$~0.13 at 330 MHz. For T$_e$=7500 K,
the corresponding upper limit to the beam averaged emission 
measure is EM $\leq$~4 $\times$~10$^4$~pc cm$^{-6}$~if the gas is at 
z=0.193 and EM $\leq$~10$^5$ pc cm$^{-6}$~if the gas is at z=0.886.
\begin{figure}
\plotfiddle{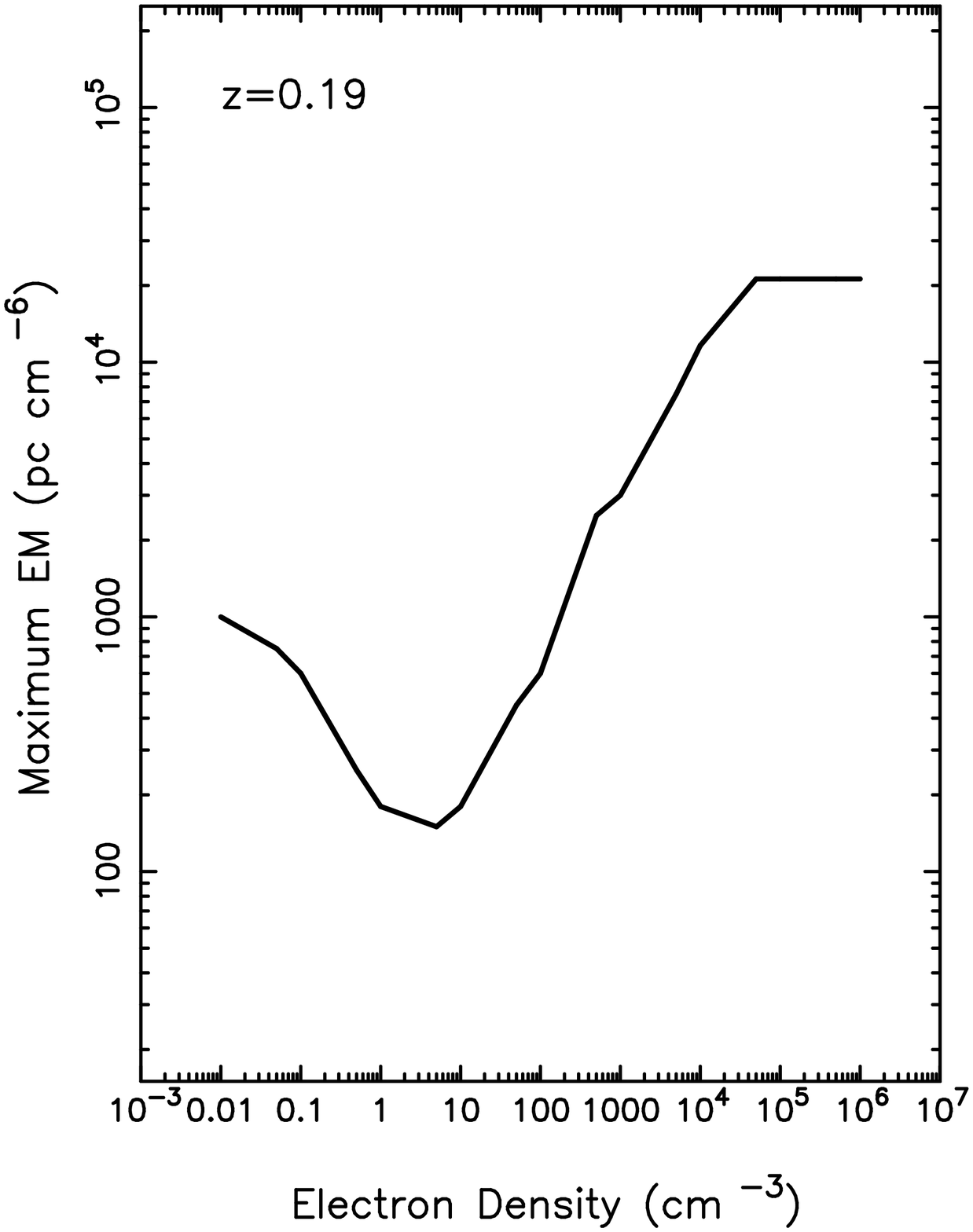}{2in}{0}{30}{20}{-180}{10}
\plotfiddle{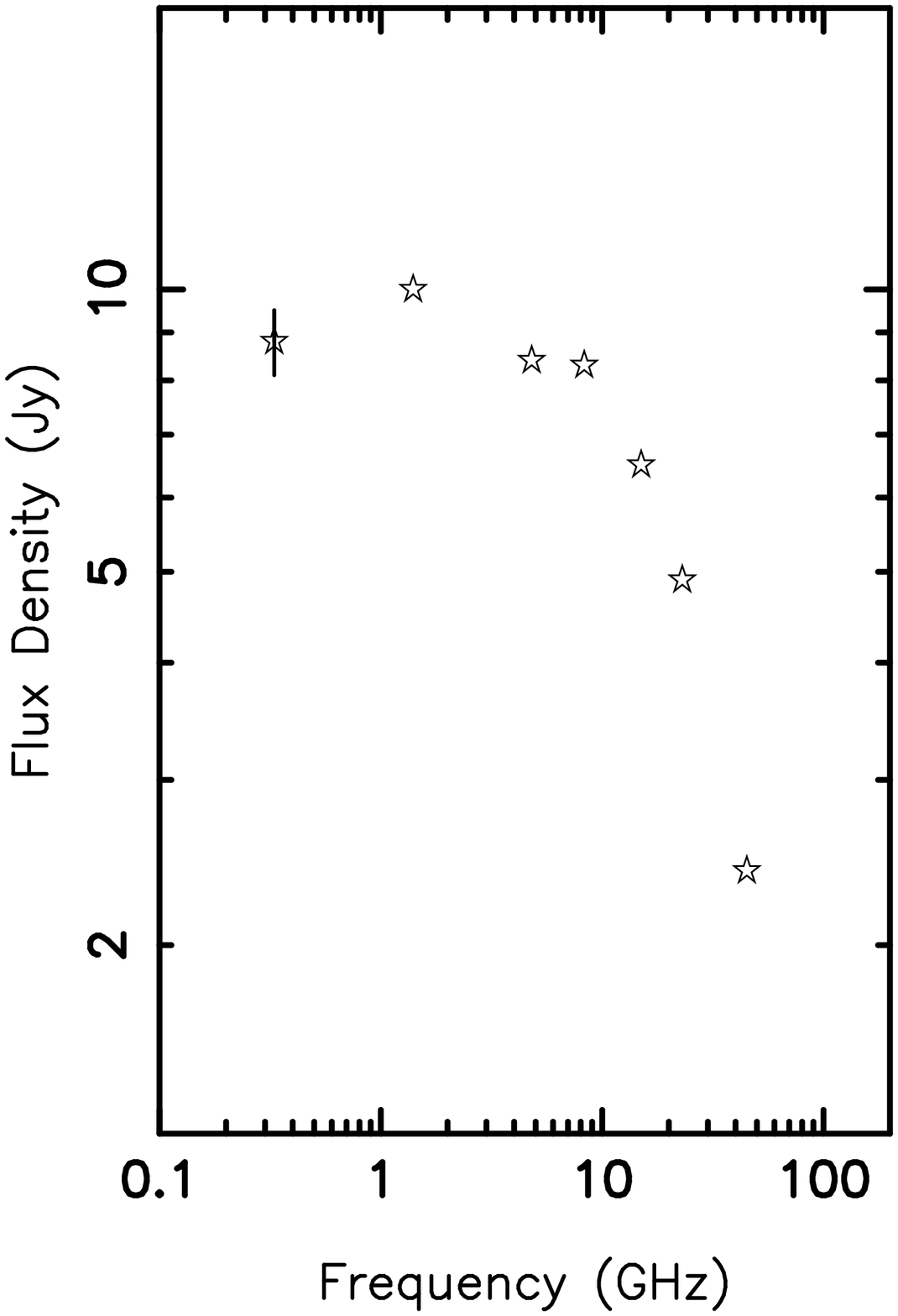}{2in}{0}{30}{20}{15}{175}
\vspace{-2.6in}
\caption{{\it Left:} The maximum allowable emission measure of the ionized gas
in the z=0.193 system as a function of its density, based on the upper
limit to the H158$\alpha$~recombination line. {\it Right:} The
measured radio continuum spectrum of PKS 1830-211}
\end{figure}
\vspace{-0.7cm}
\section{Constraints on the ionized gas at z=0.193}
\vspace{-0.3cm}
Since the HI optical depth in the z=0.193 system is the same against 
both the lensed images of the quasar (Verheijen et al. 1999), 
we assume that the line emitting gas is uniformly distributed in 
a slab against the entire continuum source. 
Figure 1 (left) shows the maximum allowable emission measure of the gas 
as a function of density that is consistent with the upper
limits to the line strength and the continuum optical depth,
assuming a line width of 80 km s$^{-1}$.

Fig 1 shows that our experiment is most sensitive to low density 
gas (n$_e$~= 1$-$10 cm$^{-3}$)
and indicates an upper limit to the {\it beam averaged} emission measure of
100 pc cm$^{-6}$~for this gas in the z=0.193 system. 
If the line emitting gas is predominantly of density 
5$-$10 cm$^{-3}$, then the size of such a region located anywhere 
in the observed beam is constrained to
be less than 200$-$300 pc in size, assuming a homogenous gas distribution. 
On the other hand, if the gas is in compact structures with 
density $\sim$10$^3$~
cm$^{-3}$, then its beam filling factor is constrained to
be less than 6 $\times$~10$^{-4}$. Higher resolution observations of
the line at comparable sensitivities and the knowledge of the system's
inclination angle will greatly improve the constraints
on the ISM in the z=0.193 absorber.
The limits for the z=0.886 system are about ten times higher.
\vspace{-0.2cm}
\acknowledgements 
We thank S. Nair for useful discussions.


\vspace{-0.7cm}
\begin{references} 
\vspace{-0.4cm}
\reference Chengalur, J.N., de Bruyn, A.G., \& Narasimha, D. 1999, \aap, 343. L79
\reference Lovell, J.E.J. et al 1996, \apj, 472, L5
\reference Verheijen, M.A.W, Carilli, C., Yun, M.S., \& Menten, K. 1999, 
private comm.
\end{references}
\end{document}